\documentclass[ aps,prd,superscriptaddress,10pt,showpacs,notitlepage,  
tightnlines,nofootinbib
]{revtex4-1}
\usepackage{bm}
\usepackage{amsmath,amssymb,amsthm}
\usepackage{latexsym,graphicx,color,subfigure}
\usepackage{enumerate}
\usepackage{amssymb}
\def\D{\mathcal{D}}
\def\half{\frac{1}{2}}
\def\Tr{ \hbox{\rm Tr}}
\def\a{\alpha}
\def\p{\partial}
\def\cfl{\textsc{cfl}}
\def\dl{\Delta_\cfl}

\def\half{\frac{1}{2}}
\def\a{\alpha}
\def\e{\epsilon}

\def\half{\frac{1}{2}}
\def\dcfl{\Delta_{\textsc{cfl}}}
\def\cfl{\textsc{cfl}}

\def \L{\textsc{l}}
\def \R{\textsc{r}}
\def \B{\textsc{b}}
\def \c{{\rm{C}}}
\def \F{{\rm{F}}}

\def\T{\tilde{T}}
\def\k{\kappa}
\begin{document}
\title{Low-energy effective worldsheet theory of a 
non-Abelian vortex in high-density QCD revisited: 
A regular gauge construction}
\author{Chandrasekhar Chatterjee}
\email{chandra@phys-h.keio.ac.jp}
\affiliation{%
Department of Physics, and Research and Education Center for Natural Sciences, Keio University, Hiyoshi 4-1-1, Yokohama, Kanagawa 223-8521, Japan
}%
\author{Muneto Nitta}%
\email{nitta@phys-h.keio.ac.jp}
\affiliation{%
Department of Physics, and Research and Education Center for Natural Sciences, Keio University, Hiyoshi 4-1-1, Yokohama, Kanagawa 223-8521, Japan
}%

\maketitle

Color symmetry is spontaneously broken 
in quark matter at high density 
as a consequence of di-quark condensations 
with exhibiting
color superconductivity. 
Non-Abelian vortices or color magnetic flux tubes  
stably exist in the color-flavor locked phase 
at asymptotically high-density. 
The effective worldsheet theory 
of a single non-Abelian vortex was previously calculated 
in the singular gauge to obtain the ${\mathbb C}P^2$ model
\cite{Eto:2009bh,Eto:2009tr}.
Here, 
we reconstruct the effective theory 
in a regular gauge without taking a singular gauge, 
confirming  the previous results in the singular gauge.
As a byproduct of our analysis, we find that non-Abelian vortices in high-density QCD do not suffer from any obstruction  for the global definition of a symmetry breaking.

\section{Introduction}

Quark matter at high temperature and/or high density 
is one of the important subjects in
both the theoretical and experimental points of view.
At high density, quark matter are expected to condensate 
by constituting Cooper pairs.
Then, color symmetry is spontaneously broken,
with exhibiting
color superconductivity. \cite{Alford:1997zt,Alford:1998mk}; 
see Refs.~\cite{Alford:2007xm,Rajagopal:2000wf} 
as a review.  The two-flavor pairing may occur in  the
 two-SC phase in which up and down quarks participate in condensations 
at intermediate density. 
At asymptotically high densities, 
if we can neglect  the strange quark mass, 
the system possesses an $SU(3)$
global flavor symmetry.
In that region, 
it may be possible to have a three-flavor pairing state, which is known as   
the ``color-flavor locked (CFL)'' phase, 
in which up,  down and strange quarks 
participate in condensations. 
The Ginzburg-Landau (GL) free energy~\cite{Giannakis:2001wz,Iida:2000ha,Iida:2001pg} shows that
 in the CFL ground state the baryon number $U(1)_{\rm B}$,
color $SU(3)_{\rm C}$, and flavor $SU(3)_{\rm F}$ symmetries
are spontaneously broken down to the diagonal subgroup 
$SU(3)_{\rm C+F}$. 
In particular, 
 $U(1)_{\rm B}$ and color symmetry breakings
 lead to superfluidity and color superconductivity, respectively.  
 Therefore, when the CFL medium rotates, $U(1)_{\rm B}$ superfluid vortices 
 with the quantized circulations  
 are created along the rotation axis 
 \cite{Iida:2000ha,Iida:2001pg,Forbes:2001gj} 
 as in the case of helium superfluids and ultracold atomic gases. 
 Compared to the quantized unit 
 circulation of $U(1)_{\rm B}$ superfluid vortices,  
 vortices with smaller circulations (1/3 quantized circulations) 
 exist, which also carry color magnetic fluxes.
They are non-Abelian vortices or color magnetic flux tubes 
\cite{Balachandran:2005ev,Nakano:2007dr,Nakano:2008dc,Eto:2009kg};
see Ref.~\cite{Eto:2013hoa} for a review.
It was conjectured that one $U(1)_{\rm B}$ superfluid vortex 
is energetically split into a set of three color flux tubes 
with total color cancelled out \cite{Nakano:2007dr},
which has been recently confirmed numerically 
\cite{Alford:2016dco}. 

One non-Abelian vortex breaks the color-flavor symmetry $SU(3)_{\rm C+F}$ further into its subgroup in the vicinity of the core, 
generating Nambu-Goldstone  modes (or collective coordinates) 
 which parametrize a complex projective space 
${\mathbb C}P^2\simeq 
SU(3)_{\rm C+F}/[SU(2) \times U(1)]$. 
These ${\mathbb C}P^2$ modes are localized 
around the vortex core 
and propagate along the vortex line
as gapless excitations 
\cite{Eto:2009bh,Eto:2009tr}. 
A lot of rich physics were obtained from 
these ${\mathbb C}P^2$ modes. 
When the coupling of the ${\mathbb C}P^2$ target space 
to electromagnetic fields is introduced
\cite{Vinci:2012mc},
it implies that 
a vortex lattice system 
behaves as a polarizer \cite{Hirono:2012ki}.
It also shows the Aharanov-Bohm scattering 
of charged particles such as electrons and muons  
\cite{Chatterjee:2015lbf}.
The quantum mechanically induced gap shows 
the confinement of monopoles 
in the CFL phase, that is, 
quark condensations lead monopole confinement 
\cite{Gorsky:2011hd,Eto:2011mk}, 
which gives  evidence of hadron-quark duality 
to the confinement phase in which quark confinement 
is expected to occur due to monopole condensations. 
Vortices interact with gluons by a topological interaction
\cite{Hirono:2010gq}, 
implying that 
in the system of multiple vortices such as 
a vortex lattice, 
${\mathbb C}P^2$ modes in individual vortices 
are aligned by the interaction,
exhibiting color ferromagnetism \cite{Kobayashi:2013axa}.
Even with such discoveries of rich physics, 
there remains one technical problem 
in the derivation of the effective Lagrangian 
in Refs.~\cite{Eto:2009bh,Eto:2009tr};
in these references a singular gauge was taken 
to construct the effective ${\mathbb C}P^2$ model.
It is well known that,  in general, 
one needs a careful treatment in the singular gauge.
For instance, in the Abelian-Higgs model 
relevant for conventional metallic superconductors,
a magnetic flux of a vortex is unphysically 
removed by taking the singular gauge.
Therefore, one needs carefully 
to check whether the results
in Refs.~\cite{Eto:2009bh,Eto:2009tr}
for the effective action are correct  
and whether or not any additional term exists.

In this paper, we construct the effective Lagrangian of a single 
non-Abelian vortex 
in a regular gauge without taking a singular gauge
and confirm that 
the result of the effective Lagrangian calculated 
in the singular gauge in Refs.~\cite{Eto:2009bh,Eto:2009tr}
is correct
and no further term exists. 
To do this, we generalize the ansatz for the gauge field 
used in the singular gauge, because 
it does not solve the Gauss-law constraint in the regular gauge.
We introduce two different profile functions 
that depend on both the radial coordinate $r$ and 
the azimuthal angle $\theta$,   
in contrast to the singular gauge for which the profile function depends only on $r$.
The profile functions of the zero modes are expanded in terms of partial waves, 
and we check the asymptotic behaviors of all the partial wave modes. 
By inserting the  solutions of the partial wave modes into the original GL action, 
we derive the effective action of the vortex 
as the ${\mathbb C}P^2$ action living on the vortex worldsheet. 
The mode previously found in the singular gauge comes out as a normalizable mode 
among the partial wave modes discussed in this paper.
By showing that the rest of the partial modes are all non-normalizable, 
we prove that the previous result on the effective theory on the vortex is correct. 

This paper is organized as follows.
In Sec.~\ref{sec:GL}, we review the GL effective theory, 
a non-Abelian vortex solution and its properties.
In Sec.~\ref{sec:regular}, we construct the effective 
theory of a single vortex in a regular gauge.
Section~\ref{sec:summary} is devoted to a summary and discussion.

\section{The Ginzburg-Landau description of dense QCD 
and a non-Abelian vortex}\label{sec:GL}

\subsection{Ginzburg-Landau effective theory}
We start with  the time-dependent GL  Lagrangian for the CFL order parameters $\Phi_\L$ and $\Phi_\R$ which are defined as 
  di-quark condensates,
\begin{eqnarray}
&&{\Phi_{\L}}_a^{\it A}  \sim  \e_{abc}\e^{\it ABC} {q_\textsc{l}}_b^{\it B} \mathcal{C}{q_\L}_c^{\it C}, 
\quad 
{\Phi_\R}_a^{\it A}  \sim  \e_{ abc}\e^{\it ABC} {q_\R}_b^{\it B} \mathcal{C}{q_\R}_c^{\it C},
\end{eqnarray}
 where $q_{\L / \R}$ stand for left- and right-handed quarks with ${a, b, c}$ as fundamental color ($SU(3)_{\c}$), ${\it A, B, C}$ as fundamental flavor ($SU(3)_{\L/\R}$) indices and $\mathcal{C}$ is the charge conjugation operator.
Since at a high-density region a perturbative calculation shows mixing terms between $\Phi_\L$ and $\Phi_\R$ are negligible, we simply assume $\Phi_\L= - \Phi_\R=\Phi$ and fix their relative phase to unity. 
 The transformation properties of the field $\Phi$ can be written as
\begin{eqnarray}
 \Phi' = e^{i\theta_\B}U_\c \Phi U_\F^{-1}, 
   \quad
 e^{i\theta_\B} \in U(1)_{\rm{B}}, 
   \quad 
 U_\c \in SU(3)_\c,
   \quad 
 U_\F \in SU(3)_\F .
\end{eqnarray}
Here $SU(3)_\F$ is defined as the diagonal subgroup ($SU(3)_{\L + \R}$)of the full flavor group $SU(3)_\L \times SU(3)_\R$. There is a redundancy  of the discrete symmetries, and the actual symmetry group is given by
\begin{eqnarray}
 G  =
    \dfrac{SU(3)_{\c} \times SU(3)_{\F} \times U(1)_{\rm{B}}}
   {\mathbb{Z}_3 \times \mathbb{Z}_3}.
\label{eq:sym_G}
\end{eqnarray}
The  Lagrangian  has been obtained as a low-energy effective theory of the high-density QCD in the CFL phase
\cite{Giannakis:2001wz,Iida:2000ha,Iida:2001pg,Abuki:2006dv}
\footnote{In this paper we are ignoring the first-order time derivative term for simplicity since it makes the vortex dyonic.}
 \begin{align}
 \mathcal L_{\rm{GL}} &=   \Tr\left[- \epsilon_3 F_{0i}F^{0i}-\frac1{2\lambda_3}F_{ij}F_{ij} 
 + K_{0}\nabla_{0} \Phi^{\dagger}\nabla_{0} \Phi-K_{3}\nabla_{i} \Phi^{\dagger}\nabla_{i} \Phi -  V(\Phi) \right] ,\nonumber\\
 V(\phi) &= - m^{2}\Phi^{\dagger}\Phi 
+ \beta \left[(\Tr[\Phi^{\dagger}\Phi])^{2} + \Tr\left\{(\Phi^{\dagger}\Phi)^{2}\right\}\right] +\frac{3 m^{4}}{16\beta}\, ,
  \label{eq:lagrangian}
\end{align}
where $F_{\mu\nu} = \p_\mu A_\nu - \p_\nu A_\mu - ig_s[A_\mu,A_\nu], \nabla_\mu = \p_\mu -i g_s A_\mu$, $\mu = 0,1,2,3$ is the space-time index,  $\{ i,j\} = \{1,2,3\}$ are spatial indices,  $\lambda_3$ is a magnetic permeability and $\epsilon_3$ is a dielectric constant for gluons. 
Here we ignore the strange quark mass.
The static GL free energy functional  can be  defined  as
 \begin{align}
{ \cal E}_{\rm{GL}} =   \Tr\left[\frac1{2\lambda_3}F_{ij}F^{ij}+ K_{3}\nabla_{i} \Phi^{\dagger}\nabla_{i} \Phi +  V(\Phi) \right] .\quad
  \label{eq:Free-energy}
\end{align}
 The coefficients in the expression above may be calculated directly from the QCD Lagrangian using perturbative techniques. We quote here the standard results obtained in the literature \cite{Giannakis:2001wz,Iida:2000ha,Iida:2001pg} through perturbative calculations in QCD as
$   \beta =   \frac{7 \zeta(3)}{8 (\pi T_{c})^{2}} N(\mu) , K_3= \frac{2}{3}\beta, K_0 = 3K_3,
m^{2}=-4 N(\mu) \log \frac{T}{T_{c}},
N(\mu)=\frac{\mu^{2}}{2\pi^{2}}  , 
g_{s}=\sqrt\frac{24 \pi^{2}\lambda}{27 \log \mu/\Lambda}, T_{c}\sim\mu \exp\left(-\frac{3 \pi^{2}}{\sqrt2 g_{s}}\right) \,$,
where $\mu$ is the chemical potential, $\Lambda$ the QCD scale and $T_{c}$ the critical temperature. 

 The vacuum expectation value of $\Phi$ can be computed by minimizing the  potential  defined at Eq.~(\ref{eq:lagrangian}) as:
\begin{eqnarray}
\left<\Phi\right>=\Delta_{\textsc{cfl}} {\bf 1}_{3},\quad \Delta_{\textsc{cfl}}^{2}\equiv\frac{m^{2}}{8\beta}\,.
\label{eq:vacuumLG}
\end{eqnarray}
In the ground state [Eq.~(\ref{eq:vacuumLG})],
the full symmetry group $ G$ is spontaneously broken down to 
$
{{H} }
= \frac{SU(3)_{\rm C+F}}{\mathbb{Z}_3}
\label{H} $
and the order parameter space  becomes
\begin{eqnarray}
{ G/H} \simeq 
{SU(3) \times U(1) \over {\mathbb Z}_3}
=U(3).
\label{G/H}
\end{eqnarray}
 Masses of gauge bosons and scalars are given by the following\cite{Eto:2009kg},
$m^{2}_{g}= g^{2}_{s}\Delta_{\textsc{cfl}}^{2}K_{3}\lambda_3, \, m^{2}_{\zeta}=\frac{2 m^{2}}{K_{3}},\, m_{\chi}^{2}=\frac{4 \lambda_{2}\Delta_{\textsc{cfl}}^{2}}{K_{3}},\, m_{\varphi}^{2}=0,$
where $\varphi$ is the massless Nambu-Goldstone boson related to the breaking of $U(1)_{\rm B}$ symmetry, and $\zeta$ and $\chi$ are, respectively, the trace and traceless part of $\Phi$.

The static equations of motion can also be directly found from the free energy in Eq.~(\ref{eq:Free-energy}), and they read as:
\begin{align}
\nabla_{i}F_{ij} & =  i g_{s}K_{3}\lambda_3 \left[  \nabla_{j}\Phi \Phi^{\dagger}-\Phi(\nabla_{j}\Phi)^{\dagger}-\frac13 \Tr\left(\nabla_{j}\Phi \Phi^{\dagger}-\Phi(\nabla_{j}\Phi)^{\dagger}\right)\right]\,, \nonumber \\
\nabla_{j}^2\Phi &= \frac{1}{K_{3}} \left[-m^{2} + 2\beta \left\{ \Phi \Phi^{\dagger} +\Tr(\Phi^{\dagger}\Phi)\right\}\right] \Phi\,.
\label{eq:motion}
\end{align}
\subsection{Non-Abelian vortex or color magnetic flux tube}
Let us first briefly review a few primary  features of the non-Abelian vortices in the CFL phase 
in the absence of the electromagnetic interaction. It can be easily noticed from Eq.~(\ref{G/H}) that $\pi_1 ({ G/H}) = \mathbb Z$. This nonzero fundamental group 
implies the existing vortices. Since the broken $U(1)_{\rm{B}}$ is a global symmetry, 
the vortices are global vortices  
or superfluid vortices 
\cite{Balachandran:2005ev}. 
The structure of these vortices can be understood by the orientation and winding of the configuration of the condensed scalar field $\Phi$ in the far away from vortex core.
We place a vortex along the $z$ direction and use the cylindrical coordinates in this paper. 
One can write down the ansatz as \cite{Balachandran:2005ev,Nakano:2007dr,Nakano:2008dc,Eto:2009kg}
 \begin{eqnarray}
 \label{colorvortexconfig1}
\Phi(r, \theta)  = 
\dcfl\left(
\begin{array}{ccc}
  e^{i\theta}f_1(r)& 0 & 0 \\
  0 &  f_2(r) & 0\\
  0 & 0 & f_2(r)
\end{array}
\right), \, 
A_i(r) = - \frac{1}{3g_s} \frac{\epsilon_{ij} x_j}{r^2} A(r) \left(
\begin{array}{ccc}
 2 & 0 & 0 \\
 0 & - 1 & 0\\
 0 & 0& -1
\end{array}
\right) , i = \{1, 2\}
\end{eqnarray} 
where $f_1 , f_2,$ and $A(r)$ are the profile functions.
The  GL free energy can be written by inserting the ansatz  into Eq.~(\ref{eq:lagrangian}) as
\begin{align}
{ \cal E}_{\rm{GL}} = 2\pi \int r dr \left[ \frac{2}{3 g_s^2\lambda_3 r^2} (\p_r A)^2 +  K_3 \dl^2\left\{(\p_r f_1)^2 + 2(\p_r f_2)^2 +  \frac{f_1^2}{9r^2}( 3 - 2A)^2 
+  2 \frac{f_2^2}{9r^2} A^2 + \dl^2 \left[  f_1^2 -  f_2^2 \right]^2 \right.\right. \nonumber \\ \left.\left.+ 2 \dl^2\left[ f_1^2 + 2 f_2^2  - 3\right]^2\right\} \right]
\end{align}
The form of the profiles  $f_1 , f_2,$ and $A(r)$ can be calculated numerically with the boundary condition,
\begin{eqnarray}
\label{vortexboundary}
 f_1(0) = 0, \quad 
\p_r f_2(r)|_0 = 0, \quad 
A(0) = 0,   \quad f_1(\infty) = f_2(\infty) = 1,  \quad 
A(\infty) = 1 .
\end{eqnarray}

The vortex configuration in 
Eq.~(\ref{colorvortexconfig1})  breaks the unbroken color-flavor diagonal 
 $SU(3)_{\rm{\c+\F}}$ symmetry as
 \begin{eqnarray}
 SU(3)_{\rm{\c+\F}}  \rightarrow SU(2)\times U(1)
\end{eqnarray}
showing the existence of degenerate solutions. 
This degeneracy is due to the existence of
 Nambu-Goldstone modes 
parametrizing a coset space
$\frac{SU(3)}{SU(2)\times U(1)} \simeq {\mathbb C}P^{2}$ \cite{Nakano:2007dr}.
The low-energy excitation and interaction of these zero modes can be calculated by
 the effective ${\mathbb C}P^{2}$ sigma model action 
\cite{Eto:2009bh}.  
Generic solutions on the ${\mathbb C}P^{2}$ space can be found by just applying a global transformation by 
a reducing matrix\cite{Delduc:1984sz},
 \begin{eqnarray}
 \label{CP2}
{U}(\bm{\phi}) = 
\frac{1}{\sqrt{X}}\left(
\begin{array}{cc}
 1 & -{\bm\phi}^\dagger \\
{\bm\phi} &  X^\half Y^{-\half} 
\end{array}
\right), \quad 
X = 1 +{\bm\phi}^\dagger{\bm\phi}, \qquad Y = {\bf 1}_3 + {\bm\phi}{\bm\phi}^\dagger, 
\end{eqnarray}
where ${\bm\phi} = \{{\bm\phi}_1, {\bm\phi}_2\}$ are inhomogeneous 
coordinates of the ${\mathbb C}P^{2}$.
The vortex solution  with a generic orientation  takes the form,
\begin{eqnarray}
 \label{colorvortexconfigcp2}
\Phi(r, \theta)  = 
\dcfl {U}({\bm\phi})\left(
\begin{array}{ccc}
  e^{i\theta}f_1(r)& 0 & 0 \\
  0 &  f_2(r) & 0\\
  0 & 0 & f_2(r)
\end{array}
\right){U^\dagger}({\bm\phi}), \,
A_i(r) = -  \frac{\epsilon_{ij}  x_j}{3g_s r^2} A(r)  {U}({\bm\phi})\left(
\begin{array}{ccc}
 2 & 0 & 0 \\
 0 & - 1 & 0\\
 0 & 0& -1
\end{array}
\right){U^\dagger}({\bm\phi}).\nonumber\\
\label{Aicp2}
\end{eqnarray}

\section{The construction of the effective action 
of a non-Abelian vortex}\label{sec:regular}

The effective action can be computed by prompting the moduli parameter ${\bm\phi}$ to fields fluctuating on the vortex worldsheet 
in the $t$-$z$ plane, 
which is known as the moduli approximation \cite{Manton:1981mp} 
(see also Refs.~\cite{Eto:2006uw,Eto:2006pg}). 
So when one inserts the 
rotated solution in Eq.~(\ref{colorvortexconfigcp2}) into the free energy, 
the static energy part is separated out from the rest. 
The other terms which would be relevant 
for small fluctuations can be written as
\begin{eqnarray}
\label{effaction1}
\mathcal{L}_{\rm eff} =  \sum_\alpha c_\alpha \Tr \left[ F_{i\alpha}F_i^{\alpha} + \k_\alpha |\D^\alpha \Phi|^2 \right] 
\end{eqnarray}
where  $\alpha = \{0,3\}$ is the worldsheet index,  $i = \{1,2\}$, $c_0 = \epsilon_3$, $c_3 = \frac{1}{\lambda_3}$ and $\k_\alpha = \frac{K_\alpha}{c_\alpha}$.  The raising and lowering of the index $\alpha$  are done
by the Minkowski signature $(+, -)$ for $\{0, 3\}$.  So the  equations of motion for zero (the Gauss's law) and the third component can be expressed as
\begin{eqnarray}
\label{EOM}
\D_iF^{i\alpha} = -i g_s \k_\a T^a \Tr \left[\Phi^\dagger T^a \D^\a \Phi - (\D^\alpha\Phi)^\dagger T^a \Phi\right].
\end{eqnarray}
These equations are generated due to the fluctuation of the zero mode along the vortex.  The ansatz for  the generated gauge fields
which solve the above Eq.~(\ref{EOM})  can be expressed as \cite{Bolognesi:2015ida}
\begin{eqnarray}
&&A_\a = \rho_\a(r, \theta) W_\a+ \eta_\a(r, \theta) V_\a, \qquad\a={0,3}
\label{Aalpha}
\end{eqnarray}
where $\rho_\alpha$ and $\eta_\a$ are profile functions which are to be determined by minimizing the action or by solving Eq.~(\ref{EOM}) and
\begin{eqnarray}
W_\a = i \p_\a \T \, \T,\qquad V_\a = \p_\alpha \T, \qquad \T = UTU^\dagger, \qquad T = \rm{diag}(1, -1, -1), 
\end{eqnarray}
where $U(\bm{\phi})$ is defined in the last section Eq.~(\ref{CP2}). These satisfy the commutation relations:
$ [W_\a, \T] = 2iV_\a,\, [V_\a, \T]= -2i W_\a,\, \Tr W_\a W^\a = \Tr V_\a V^\a,\, \Tr W_\a V^\a = 0$.
Here, we can see that $W_\a$ and $V_\a$ are orthogonal to each other.  $W_\a$ and $V_\a$ are also orthogonal to the direction of $A_i$ defined in Eq.~(\ref{Aicp2}). We can see this if we expand all three
matrices as
\begin{eqnarray}
  {U}({\bm\phi})\frac{1}{3}\left(
\begin{array}{ccc}
 2 & 0 & 0 \\
 0 & - 1 & 0\\
 0 & 0& -1
\end{array}
\right){U^\dagger}({\bm\phi}) =\frac{1}{6} \bm{1} + \half  \tilde{T} ,\,\, W_\alpha = i[\p_\a U U^\dagger - \tilde{T}\p_\a U U^\dagger \tilde{T}], \,\,V_\a = -i[i \p_\a U U^\dagger,  \tilde{T} ].
\label{delduc}
\end{eqnarray}
 $W_\alpha$ is the Delduc-Valent projection on ${\mathbb C}P^2$ \cite{Delduc:1984sz} and was used in singular gauge computation. Here we introduce $V_\alpha$ as a new component in the ansatz (\ref{Aalpha}) to solve  Gauss's law, getting 
$A_\a$  orthogonal  to the $A_i$ direction  indicating the fluctuation of the Nambu-Goldstone mode in the orthogonal direction of the background field, which is   true because the Nambu-Goldstone bosons are generated by broken generators.

To compute the effective action, we have to insert the ansatz in Eq.~(\ref{Aalpha}) into the action in Eq.~(\ref{effaction1}). Before we do so, let us compute the field strength of the gauge field and matter coupling separately. 
The first term of the field strength($ F_{i\a} = \p_i A_\a - \D_\a A_i$) can be written as
\begin{align}
&& \p_i A_\a  = \left(\frac{x_i}{r}\p_r \rho_\a(r, \theta) - \frac{\epsilon_{ij}x^j}{r^2}\p_\theta \rho_\a(r, \theta)\right) W_\a + \left( \frac{x_i}{r}\p_r\eta_\a(r, \theta) - \frac{\epsilon_{ij}x^j}{r^2}\p_\theta \eta_\a(r, \theta)\right) V_\a.
\end{align}
The second term  becomes
\begin{eqnarray}
\D_\a A_i &=& \p_\a A_i - i g_s[A_\a, A_i]
 = \frac{\epsilon_{ijx^j}}{ r^2}A(r)\left\{\eta_\a(r, \theta) W_\a - \sigma_\a(r, \theta) V_\alpha \right\}
\end{eqnarray}
where we have defined  $2 g_s \sigma_\a = 1 + 2 g_s \rho_\a$.
So, we can insert  the field strength, 
\begin{eqnarray}
 F_{i\a} &&=  \p_i A_\a - \D_\a A_i \nonumber \\
 && = \left(\frac{x_i}{r}\p_r \rho_\a - \frac{\epsilon_{ij}x^j}{r^2}(\p_\theta \rho_\a + A\eta_\a)\right) W_\a + \left( \frac{x_i}{r}\p_r\eta_\a  - \frac{\epsilon_{ij}x^j}{r^2}(\p_\theta \eta_\a - A \sigma) \right) V_\a 
\end{eqnarray}
into the  kinetic term of the gauge field in Eq.~(\ref{effaction1})  to yield
\begin{align}
\label{Fsquare}
 \Tr F_{i\a} F_{i}^{\a} =  \left[(\p_r \rho_\a)^2 + (\p_r \eta_\a)^2 + \frac{1}{r^2} (\p_\theta \rho_\a + A\eta_\a)^2 + \frac{1}{r^2} (\p_\theta \eta_\a - A \sigma_\a)^2 \right] \Tr V_\a V^\a.
\end{align}

From the covariant derivative of the matter field $\Phi$,
\begin{align}
\D_\a \Phi =&\,\, \p_\a \Phi - i g_s A_\a \Phi\nonumber\\
 =&\, \dl\left[  \frac{f_1 e^{i\theta} - f_2}{2}(1 +  g_s \rho_\a ) -i g_s \eta_\a  \frac{f_1 e^{i\theta} + f_2}{2}  \right] V_\alpha - g_s\dl\left(\eta_\a \frac{f_1 e^{i\theta} - f_2}{2}  + i \rho_\a  \frac{f_1 e^{i\theta} + f_2}{2} \right) W_\a, 
\end{align}
we compute the $|\D_\a \Phi |^2 $ as
\begin{align}
\frac{|\D_\a \Phi |^2 }{\dl^2} =  
 \frac{1}{4}\left[(1 +  2g_s \rho_\a +2g_s^2 \rho^2_\a )(f_1^2 + f_2^2) + 2  g_s^2 \eta^2_\a (f_1^2 + f_2^2) - (1 +  2g_s \rho_\a )2 f_1f_2 \cos\theta\right.\nonumber\\\left. - 4 g_s f_1f_2 \eta_\a  \sin\theta\right]  \Tr V_\a V_\a .
 \label{dphi2}
\end{align}
By changing the variables 
from  $\rho_\a$ to $\sigma_\a = \frac{ 1 + 2 g_s \rho_\a}{2 g_s}$, 
 Eq.~(\ref{dphi2}) can be simplified as 
\begin{eqnarray}
\label{Dphisquare}
\frac{4|\D_\a \Phi |^2 }{\dl^2}= \left[\half (f_1^2 + f_2^2) +  2 g_s^2 \left(\sigma_\a^2+ \eta_\a^2\right)(f_1^2 + f_2^2)  - 4 g_s f_1f_2\left(\sigma_\a  \cos\theta     +  \eta_\a  \sin\theta\right)\right] \Tr V_\a V_\a.
\end{eqnarray}
 Let us  define here a complex  scalar field as 
\begin{eqnarray}
\Psi_\a(r, \theta) = \sigma_\a(r, \theta) + i \eta_\a(r,\theta)
\end{eqnarray}
and we can rewrite  the action in terms of these fields. 
Equation~(\ref{Fsquare}) becomes 
\begin{align}
 \Tr F_{i\a} F_{i}^{\a} 
= \sum_\a \left[|\p_r \Psi_\a|^2 +  \frac{1}{r^2} |\D_\theta \Psi_\a|^2 \right] \Tr W_\a W^\a
\end{align}
where $\D_\theta = \p_\theta - i A(r)$. 
Equation~(\ref{Dphisquare}) can also be rewritten as
\begin{align}
\frac{|\D_\a \Phi |^2 }{\dl^2} 
= \frac{1}{4}\left[\half (f_1^2 + f_2^2) +  2 g_s^2 |\Psi_\a|^2(f_1^2 + f_2^2)      - 2 g_s f_1f_2 (\Psi_\a e^{-i\theta} + \Psi_\a^* e^{i\theta}) \right] \Tr W_\a W^\a .
\end{align}
The effective Lagrangian in Eq.~(\ref{effaction1}) can be written as
\begin{align}
\mathcal{L}_{\rm eff} = \sum_\a c_\alpha \int d^2 x\,\, \Tr \left[  F_{i\alpha}F_i^{\alpha} + \k_\alpha |\D^\alpha \Phi|^2 \right] 
= \sum_\a \mathcal{I}_\alpha \,\, \Tr W_\a W^\a
\label{I}
\end{align}
where $\mathcal{I}_\alpha$ are the coefficients of the ${\mathbb C}P^2$ action,  defined by
\begin{align}
\mathcal{I}_\alpha = c_\alpha \int r dr d\theta\,\, \left[|\p_r \Psi_\a|^2 +  \frac{1}{r^2} |\D_\theta \Psi_\a|^2 +
\frac{\dl^2\k_\a}{4}\left\{\half (f_1^2 + f_2^2) +  2 g_s^2 |\Psi_\a|^2(f_1^2 + f_2^2)      - 2 g_s f_1f_2 (\Psi_\a e^{-i\theta} + \Psi^*_\a e^{i\theta}) \right\}\right] .
\end{align}
The effective Lagrangian can be written explicitly as 
the form of the ${\mathbb C}P^2$ model,
\begin{eqnarray}
{\cal L}_{\rm eff} = \mathcal{I}_0 \left\{{\dot{\bm{n}}}^\dagger\dot{\bm{n}} + ({\bm{n}}^\dagger\dot{\bm{n}})({\bm{n}}^\dagger\dot{\bm{n}})\right\} - \mathcal{I}_3 \left\{{\p_z{\bm{n}}}^\dagger\p_z{\bm{n}} + ({\bm{n}}^\dagger\p_z{\bm{n}})({\bm{n}}^\dagger\p_z{\bm{n}})\right\} 
\end{eqnarray}
where $\bm{n}$ are the homogeneous coordinates of the ${\mathbb C}P^2$ space, which can be written in terms of 
the inhomogeneous coordinates as 
\begin{eqnarray}
\bm{n} = \frac{1}{\sqrt{1 + \bm{\phi}^\dagger \bm{\phi}}}
\left(
\begin{array}{c}
  1  \\ \bm{\phi}    
\end{array}
\right).
\end{eqnarray}
If we rescale the  $z$ coordinate as
\begin{eqnarray}
z^\prime = \sqrt{\frac{\mathcal{I}_0}{\mathcal{I}_3}} \,\,\,z
\end{eqnarray}
then the effective Lagrangian becomes 
\begin{eqnarray}
{\cal L}_{\rm eff} = \mathcal{I}_0  \left[{\p_\alpha{\bm{n}}}^\dagger\p^\alpha{\bm{n}} + ({\bm{n}}^\dagger\p_\alpha{\bm{n}})({\bm{n}}^\dagger\p^\alpha{\bm{n}})\right].
\end{eqnarray}
By expanding the field $\Psi_0$ in terms of partial waves as 
\begin{eqnarray}
\Psi_0 = \sum_m \Psi_m(r) e^{im\theta},
\end{eqnarray}
$\mathcal{I}_0$ can be written in terms of the partial wave modes as
\begin{align}
\mathcal{I}_0 = c_0\sum_m \int r dr d\theta\,\, \left[(\p_r \Psi_m)^2 +  \frac{(m -A(r))^2}{r^2}  \Psi_m^2 
+ \frac{\dl^2\k_0}{4}\left\{\half (f_1^2 + f_2^2) +  2 g_s^2 \Psi_m^2(f_1^2 + f_2^2)   \right.\right. \nonumber\\  \left.\left. -  4 g_s f_1f_2 \Psi_m\cos(m-1)\theta
\right.\Big\}\right.\Big] .\label{I2}
\end{align}
 It is easy to check that the last term of Eq.~(\ref{I2}) 
vanishes after the theta integration unless $m = 1$.  So we write $\mathcal{I}_0$ as
\begin{align}
\mathcal{I}_0 = \sum_m 2\pi c_0 \int r dr \,\, \left[(\p_r \Psi_m)^2 +  \frac{(m -A(r))^2}{r^2}  \Psi_m^2 +
\frac{\dl^2\k_0}{4}\left\{\half (f_1^2 + f_2^2) +  2 g_s^2 \Psi_m^2(f_1^2 + f_2^2)      -  4 g_s f_1f_2 \Psi_m \delta_{m1}
\right\}\right] .\label{I0}
\end{align}
We minimize the  above 
integral by solving the equations for the modes,
\begin{align}
\frac{1}{ r} \p_r r \p_r \Psi_1 - \frac{(1 - A(r))^2}{ r^2}  \Psi_1  &= \frac{\dl^2\k_0 g_s}{2} \left[(f_1^2 + f_2^2)g_s \Psi_1 - f_1f_2\right], \label{norm1}\\
 \frac{1}{ r}\p_r r \p_r \Psi_m - \frac{(m - A(r))^2}{ r^2}  \Psi_m  &=  \frac{\dl^2\k_0 g_s^2}{2} (f_1^2 + f_2^2) \Psi_m, \, \, m \ne1 .\label{nnorm}
\end{align}
These equations can  also  be
derived directly from the equations of motion in Eq.~(\ref{EOM}).

One should notice that 
Eq.~(\ref{norm1}) for $m=1$ was derived in the singular gauge in Ref.~\cite{Eto:2009bh} (for $K_0 = K_3$) 
but the rest, where $m\neq1$ were absent in the singular gauge.
The solution of Eq.~(\ref{norm1}) is normalizable ($m=1$) and can be solved \cite{Eto:2009bh} numerically with the boundary condition
$\Psi_1(0)=0$ and $\Psi_1(\infty) =  \frac{1}{2g_s}  $. 
Large distance behavior  of the solution can be expressed as 
\begin{eqnarray}
\Psi_1 \rightarrow_{r \rightarrow \infty} \frac{1}{2g_s} - \frac{1}{\sqrt \xi} e^{ -\xi}
\end{eqnarray}
and it is easy to show that the large distance value of   $\Psi_1 =  \frac{1}{2g_s} $ transforms the ansatz $A_\alpha$ in Eq.~(\ref{Aalpha}) into a pure gauge form as
\begin{eqnarray}
A_\alpha = \frac{i}{g_s} \bm{g}^\dagger  \p_\alpha\bm{g}, \quad\text{where} \quad\bm{g} = e^{-i\theta T_8^*}, \quad T_8^* = \frac{1}{3}U(\bm{\phi})\left(
\begin{array}{ccc}
 2 & 0 & 0 \\
 0 & - 1 & 0\\
 0 & 0& -1
\end{array}
\right)U^\dagger(\bm{\phi}).
\end{eqnarray}
The coefficient $\mathcal{I}_0(m=1)$ can be written in terms of the solution of the 
equations of motion derived in the above:
\begin{eqnarray}
\mathcal{I}(m =1) =\frac{2\pi m_0^2c_0}{8g_s^2} \int  r dr \,\, \left[ f_1^2 + f_2^2- 4 f_1f_2 g_s\Psi_1\right] ,
\end{eqnarray}
where $m_0^2 = g_s^2\dl^2 \k_0 $. The integrand vanishes at large distances, so the integral is finite. 

        All other modes ($m\ne1$)  are non-normalizable divergent modes. In this work, we do not solve these equations numerically. However, one can understand the behavior  
of the solutions once we analyze the asymptotic forms of the solutions.  For the case $m\ne 1$, Eq.~(\ref{nnorm}) can be expressed as
\begin{align}
\label{EOMPsi}
 \frac{1}{ r}\p_r r \p_r \Psi_m  &= \left[ \frac{(m - A(r))^2}{ r^2}  +  \frac{\dl^2\k_0 g_s^2}{2} (f_1^2 + f_2^2)  \right]\Psi_m .
\end{align}
The right-hand side of the above equation is positive definite 
(by assuming $\Psi_m$ is positive). So the solution cannot have 
a local maximum.  At large distances,
the equation becomes the modified Bessel equation,
 \begin{eqnarray}
 \xi^2 \Psi_m''+ \xi \Psi_m' - \left[(m - 1)^2   +  \xi^2\right]\Psi_m = 0
\end{eqnarray}
where $\xi = m_0 \,r$. The solution is $\Psi_m \sim \frac{1}{\sqrt \xi} e^{\pm \xi}$. 

   At small distances near the origin ($r=0$), where $f_1(r) = 0$, Eq.~(\ref{EOMPsi}) reduces to
  \begin{eqnarray}
 \left(\p_\xi^2 + \frac{1}{\xi}\p_\xi \right)\Psi_m  =   {m^2 \over \xi^2}\Psi_m.
\end{eqnarray}
The solution  is $\Psi_m \sim \xi^{\pm m}$ for $m\ne 0,1$. So $\Psi_m(0) = 0$ for $m\ne0,1$ and we may conclude that the solution diverges as $\frac{1}{\sqrt \xi} e^{ \xi}$ at large distances since it does not have any local maximum. For $m=0$, near the origin the solution could be written as either  $\log\xi$ or $C_0 + \xi^2 C_2 $, where $C_0$ and $C_2$ are constants and $C_2$ is found to be positive. So the solution diverges at large distances even if we set the value as constant at the origin.

  \section{Summary and Discussion}\label{sec:summary}

In this paper, we have analyzed the orientational ${\mathbb C}P^2$ zero modes 
of a single non-Abelian vortex in high-density QCD. 
To do so, we have followed the  standard procedure of zero-mode analysis and have written 
the effective action, as was done in the singular gauge case.
In order to solve the Gauss's law constraint in the regular gauge, we have generalized 
the ansatz of the gauge field (used in the singular gauge) 
to $A_\alpha = \rho(r, \theta) W_\alpha + \eta(r, \theta) V_\alpha$ 
by introducing a profile function $\eta$ together 
with a matrix $V_\alpha$ orthogonal to $W_\alpha$, 
neither of which  exist in the singular gauge. 
In the regular gauge, the two profile functions ($\rho$, $\eta$ ) do not only 
depend on the radial coordinate $r$ but also on the azimuthal angle $\theta$. 
These two profile functions can be combined to the real and imaginary components of a single complex profile function $\Psi(r,\theta)$.  The insertion of $A_\alpha$ together with vortex profile functions into the action gives the ${\mathbb C}P^2$ effective action on the $t$-$z$ plane 
with a front coefficient depending on the complex profile function $\Psi(r,\theta)$. The front coefficient has been determined by inserting $\Psi(r,\theta)$ after solving the equations of motion. 
We have expanded the complex profile function $\Psi$ in the partial wave basis as $\Psi = \sum_m \Psi_m(r) e^{im\theta}$ and have analyzed the asymptotic  behaviors 
of all the modes. We have found that only one mode $\Psi_1$ is normalizable,
which is identical to what was found in the singular gauge analysis. 
We have shown that all the other modes diverge exponentially at large distances. 
 Finally, we have concluded that  our regular gauge analysis established the previously known result of the existence  of   normalizable zero mode derived in singular gauge 
 and that the previously constructed effective ${\mathbb C}P^2$ Lagrangian of the single 
 vortex is correct.  
  
       Here let us discuss some points which may shed light on  the interesting features of the regular gauge. In this analysis, we have introduced a complex profile function which does not depend only on $r$ 
 but also on the azimuthal angle $\theta$. This azimuthal angle dependence of a zero mode makes the system complicated. 
 The ${\mathbb C}P^2$ Nambu-Goldstone zero modes arise as a consequence of 
 the unbroken color-flavor group $SU(3)_{\rm C+F}$ in the bulk, 
 further broken as $SU(3)_{\rm C+F}\rightarrow U(1) \times SU(2)$ inside the vortex core.
The  system restores $SU(3)_{\rm C+F}$ symmetry  at large distances from the vortex core,
 where the unbroken $SU(3)_{\rm C+F}$ group elements commute with the order parameter. 
 However,  the presence of vortex makes the order-parameter position dependent at large distances. So the embedding of color-flavor diagonal group  $SU(3)_{\rm C+F}$ inside the original symmetry group becomes space dependent. The generators of the  $SU(3)_{\rm C+F}$ changes along a path around the vortex by the action of holonomy.   The space dependence  is true only for the few generators which belong to the ${\mathbb C}P^2$ subspace of $SU(3)$ and others remain unaffected.  The azimuthal angle dependence of ${\mathbb C}P^2$ generators actually makes the zero-mode analysis complicated. So when we fluctuate the zero modes, the generated gauge field $A_\alpha$   depends on the angle  in a  complicated way.  In general, the generators may not return back to their own after a complete rotation along an encircled path around the vortex.
 In this case, it is said that there is the so-called obstruction, 
 for which 
 the profile functions in general diverges as $r^{c}$ at large distances with a constant $c$ \cite{Alford:1990mk, Alford:1990ur,Alford:1992yx,Bolognesi:2015ida}. 
 In our case, 
 after expanding our complex profile function in the partial wave basis,  we have found that the  profile functions corresponding to different partial wave modes diverge exponentially except for one normalizable mode ($m=1$).  
 Therefore, as a byproduct of our analysis, we have shown that 
 non-Abelian vortices in high-density QCD do not suffer from any obstruction.
 
 There is an alternative way to show the absence of normalizable modes other than 
 the ${\mathbb C}P^2$ modes and translational modes, 
 that is, the index theorem.
 Fermionic zero modes around 
 a single non-Abelian vortex \cite{Yasui:2010yw,Chatterjee:2016ykq}
 were studied by the index theorem
applied to the Bogoliubov-de Gennes equation 
\cite{Fujiwara:2011za}. 
The index theorem applied to bosonic modes 
should be studied in the framework of the GL theory.

\section*{Acknowledgement}
We would like to thank Mark Alford for bringing our attention to a possible problem of the singular gauge in constructing the effective action.
This work is supported by the Ministry of Education, Culture, Sports, Science (MEXT)-Supported Program for the Strategic Research Foundation at Private Universities ``Topological Science'' (Grant No. S1511006). 
C.~C. acknowledges support as an International Research Fellow of the Japan Society for the Promotion of Science (JSPS). 
The work of M.~N. is supported in part by a
JSPS Grant-in-Aid for Scientific Research (KAKENHI Grant No. 16H03984)
and by a Grant-in-Aid for
Scientific Research on Innovative Areas ``Topological Materials
Science'' (KAKENHI Grant No.~15H05855) and ``Nuclear Matter in Neutron
Stars Investigated by Experiments and Astronomical Observations''
(KAKENHI Grant No.~15H00841) from the the Ministry of Education,
Culture, Sports, Science (MEXT) of Japan.

\end{document}